\documentclass[
aps,%
12pt,%
final,%
notitlepage,%
oneside,%
onecolumn,%
nobibnotes,%
nofootinbib,%
superscriptaddress,%
noshowpacs,%
centertags]%
{revtex4}

\usepackage{graphicx}

\def\jrn#1#2#3#4{{#1} {\bf #2} (#4) #3}
\def\PRL{Phys.Rev.Lett.}
\def\PLB{Phys.Lett.B}
\def\PRD{Phys.Rev.D}

\def\YaF{Yad.Fiz.}

\begin{document} 
 
\title{Using the $e^{\pm}\mu^{\mp}~+~E^{miss}_T$ signature\\ 
          in the search for supersymmetry and lepton flavour\\
          violation in neutralino decays}
 
\author{\firstname{Yu.M.}~\surname{Andreev}} 
\email{Iouri.Andreev@cern.ch}
\affiliation{Institute for Nuclear Research RAS, Moscow, 117312, Russia}
\author{\firstname{S.I.}~\surname{Bityukov}} 
\email{Serguei.Bitioukov@cern.ch}
\affiliation{Institute for High Energy Physics, Protvino, 142281, Russia}
\author{\firstname{N.V.}~\surname{Krasnikov}}
\email{Nikolai.Krasnikov@cern.ch}
\affiliation{Institute for Nuclear Research RAS, Moscow, 117312, Russia}
\author{\firstname{A.N.}~\surname{Toropin}}
\email{Alexander.Toropin@cern.ch}
\affiliation{Institute for Nuclear Research RAS, Moscow, 117312, Russia}

\begin{abstract}
The LHC (CMS) discovery potential of the $e^{\pm}\mu^{\mp} ~+~
E^{miss}_T$ signature in the search for supersymmetry and lepton flavour 
violation in neutralino decays is studied. A detailed study is done for 
the CMS test points LM1-LM9. It is shown that for the point LM1 it is 
possible to detect lepton flavour violation in neutralino decays with 
lepton flavour violating branching $Br(\tilde{\chi}^0_2 \rightarrow 
\mu^{\pm}e^{\mp}\tilde{\chi}^0_1) \geq 0.04 Br(\tilde{\chi}^0_2 \rightarrow 
 e^{+}e^{-}\tilde{\chi}^0_1,  \mu^{+}\mu^{-} \tilde{\chi}^0_1)$
for an integral luminosity $10 fb^{-1}$. A discovery potential 
in the mSUGRA-SUSY scenario with 
$\tan \beta = 10, ~sign(\mu) = +$ ~~in the
$(m_0, ~m_{1/2})$ plane using the $e^{\pm}\mu^{\mp} ~+~ E^{miss}_T$ 
signature is determined.
\end{abstract}
 
\maketitle

\section{
\label{Intr}
                        Introduction
}

One of the goals of the Large Hadron Collider (LHC) 
\cite{Kras2004} is the discovery of supersymmetry (SUSY). 
The squark and gluino decays produce missing transverse energy from 
lightest stable superparticle (LSP) plus multiple jets 
and isolated leptons \cite{Kras2004}. One of the most interesting 
and widely discussed
signatures for SUSY discovery at the LHC is the 
signature with two opposite charge and the same flavour leptons \cite{Baer}:
$l^+l^- ~+ ~E^{miss}_T$. 
The main reason of such interest is that neutralino 
decays into leptons and LSP $\tilde{\chi}^0_2 \rightarrow l^+l^- \tilde{\chi}^0_1$ 
contribute to this signature and the distribution 
of the $l^+l^-$ invariant mass $m_{inv}(l^+l^-)$ has  the edge structure 
\cite{Abdu} that allows to determine some combination of the SUSY masses. 

The signature $e^{\pm}\mu^{\mp} ~+ ~E^{miss}_T$, can also be produced when the 
$\chi^0_2$ decays into $\tau$ pair which is only relevant at large 
$\tan \beta$  \cite{Daniel}.
Also as it is shown in Ref. \cite{Abdu} at the level
of CMSJET \cite{cmsjet} simulation the use of 
$e^{\pm}\mu^{\mp}~+~E^{miss}_T$
signature for large $\tan \beta$ allows to obtain nontrivial information
on parameters of the decay $\tilde{\chi}^0_2 \rightarrow \tilde{\tau}\tau
 \rightarrow \tau\tau\tilde{\chi}^0_1  
\rightarrow e^{\pm}\mu^{\mp}\tilde{\chi}^0_1\nu\nu\bar{\nu}\bar{\nu}$. 
 On the 
other hand, the $e^{\pm}\mu^{\mp} ~+ ~E^{miss}_T$ (with an arbitrary number 
of jets), can be used for the detection of lepton flavour violation 
in slepton decays \cite{Kras1994}-\cite{Bartl} at the LHC.

In the Minimal Supersymmetric Model (MSSM) \cite{Msugra} supersymmetry 
 is broken at some high scale $M$ by generic 
soft terms, so in general all soft SUSY breaking terms are arbitrary 
which complicates the analysis and spoils the predictive power of the theory. 
In the Minimal Supergravity Model (mSUGRA) \cite{Msugra} 
the universality of the different soft parameters 
at the Grand Unified Theory (GUT) scale 
$M_{GUT} \approx 2 \cdot 10^{16}$ ~GeV is postulated. 
Namely, all the spin zero 
particle masses (squarks, sleptons, higgses) are postulated to be equal to 
the universal value $m_0$ at the GUT scale. All gaugino particle masses 
are postulated to be equal to the universal value $m_{1/2}$ at GUT scale. 
Also the coefficients in front of quadratic and cubic SUSY soft breaking  
terms are postulated to be equal. The renormalization group equations are 
used to relate GUT and electroweak scales. The equations for the 
determination of a nontrivial minimum of the electroweak potential are 
used to decrease the number of the unknown parameters by 2. 
So the mSUGRA model depends on five unknown parameters. At present,
 the more or 
less standard choice of free parameters in the mSUGRA model includes 
$m_0, m_{1/2}, \tan \beta , A$ and $ sign(\mu) $ \cite{Msugra}. 
All sparticle masses depend on these parameters.

The goal of this work is to search for the possibility to detect SUSY 
and lepton flavour violation (LFV) using 
the $e^{\pm}\mu^{\mp} ~+ ~E^{miss}_T$ signature 
at the LHC for the Compact Muon Solenoid (CMS) detector at the level of 
full detector simulation. 
For specific calculations the mSUGRA model is used. 

The organization of the paper is the following. Section {\ref{Tede}  describes 
some useful technical details of performed simulations. In Section {\ref{Basi}
the backgrounds and cuts used to suppress the backgrounds are discussed. 
Section \ref{Resca} contains  the results of numerical calculations concerning 
the possibility to detect SUSY using the $e^{\pm}\mu^{\mp}~+~E_T^{miss}$ 
signature. In Section \ref{Selefl} the prospects of 
the detection lepton flavour violation in the neutralino decays is studied. 
In Section \ref{Insys} the influence of the systematic uncertainties on the 
value of the signal significance is discussed. 
Section \ref{Conc} contains concluding remarks.

\section{
\label{Tede}
                        Simulation details
}

The coupling constants and cross sections in the leading order (LO) 
approximation for SUSY processes and backgrounds were calculated with 
ISASUGRA 7.69 \cite{ISASUGRA}, PYTHIA 6.227 \cite{Pythia} and 
CompHEP 4.2pl \cite{compHEP}. For the calculation of the 
next-to-leading order (NLO) corrections to the SUSY cross sections 
the PROSPINO \cite{PROSPINO} code was used. For considered signal events 
and backgrounds the NLO corrections are known and 
the values of NLO cross sections (or $k$-factors) were used
for normalization of the numerical results. 

Official datasets (DST) production was used for 
the  study of CMS SUSY test points \cite{LMPoints,PTDR2} 
(LM1, LM4, LM5, LM9), lepton flavour 
violation for the point LM1
and of backgrounds (${\rm t \bar t}$, ZZ, WW, Wt, Z${\rm b \bar b}$, 
DY2$\rm \tau$).  The 
ISASUGRA 7.69 + PYTHIA 6.225 codes were used in official production.
The full detector simulation was made with 
OSCAR\_2\_4\_5 or  OSCAR\_3\_6\_0 \cite{CMSSOFT} codes. 
Digitization was made with 
ORCA\_7\_6\_1, ORCA\_8\_5\_0 or ORCA\_8\_7\_1 \cite{CMSSOFT} codes.
  
For other CMS SUSY test points (LM2, LM3, LM6, LM7, LM8) and  WZ background 
the events were generated with ISASUGRA 7.69 + PYTHIA 6.227 codes and 
CMKIN\_4\_3\_1 \cite{CMSSOFT} was used as an interface program. 
The detector simulation and hits production for the test points
(LM2, LM3, LM6) and  WZ background were made with OSCAR\_3\_6\_5 
and for digitization ORCA\_8\_7\_3 was used.
To study the two test points LM7 and LM8, background Z+jet and to prepare 
the CMS discovery plot, the CMS fast simulation program 
$FAMOS\_1\_4\_0$ \cite{CMSSOFT} was used.
 
The pile-up for the signal events are not taken into account,
but backgrounds in DSTs were produced with pile-up corresponding to 
$2 \times 10^{33} cm^{-2}s^{-1}$ luminosity.

The reconstructed electrons and muons were passed through packages defining 
lepton isolation criteria. For each electron and muon the following 
parameters are defined:
\begin{itemize}
\item{$TrackIsolation$ is a number of additional tracks with $p_T >$ 2~GeV/$c$ 
  inside a cone with 
$R \equiv \sqrt{\Delta \eta^{2} + \Delta \Phi^{2}} < $ 0.3} around the lepton.
\item{$CaloIsolation$ is a ratio of energy deposited
 in  the calorimeters (electromagnetic (ECAL) + hadronic (HCAL)) inside a cone 
with $ R = 0.13 $ around given track to the energy 
deposited inside a cone with $ R = 0.3 $.}
\item{$HEratio$ is defined as a ratio of energy deposited 
in the HCAL inside a cone with $R = 0.13$ 
to the energy deposited in the ECAL inside the same cone.} 
\item{$EPratio$ is a  ratio of 
energy deposited in the ECAL inside a cone 
with $R = 0.13$  to the momentum of the reconstructed track.}
\end{itemize} 

Data with reconstructed electrons, muons, jets and missing energy 
were stored into ROOT \cite{ROOT} files for the final analysis. 

The official datasets used for these analysis were processed 
with CRAB \cite{CRAB}.

\section{
\label{Basi}
                   Signal selection and backgrounds 
}

The SUSY production $ pp \rightarrow \tilde{q}\tilde{q}^{'}, \tilde{g}
\tilde{g}, \tilde{q}\tilde{g}$ with subsequent decays
\begin{equation}
\tilde{q} \rightarrow q^{'}\tilde{\chi}^{\pm}_{1,2}  \\,
\end{equation}
\begin{equation}
\tilde{g} \rightarrow q \bar{q}^{'} \tilde{\chi}^{\pm}_{1,2}\\,
\end{equation}
\begin{equation}
\tilde{\chi}^{+}_{1,2} \rightarrow \tilde{\chi}^0_1 e^+(\mu^{-})\nu \\,
\end{equation}
\begin{equation}
\tilde{\chi}^{-}_{1,2} \rightarrow \tilde{\chi}^0_1\mu^{-}(e^{-})\nu \\,
\end{equation}
leads to the event topology $e^{\pm}\mu^{\mp} ~+ ~E^{miss}_T$. Note that 
in the MSSM with lepton flavour conservation neutralino decays into leptons 
$\tilde{\chi}^0_{2,3,4} \rightarrow l^+l^- \tilde{\chi}^0_1$ 
($l \equiv e,\mu$) do not 
contribute into this signature and contribute only to the 
$l^{+}l^{-}~+~E^{miss}_T$ signature.
 The main backgrounds which contribute to the $e^{\pm}\mu^{\mp}$ events 
are: ${\rm t \bar t}$, WW, WZ, ZZ, Wt, Z${\rm b \bar b}$, DY2$\tau$ and Z+jet. 
It is found that ${\rm t \bar t}$ is the largest background and it
gives more than 50\% contribution to the total background. 
The following NLO values for the main background cross 
sections \cite{Slab, Zjet} are used (Table \ref{Cross}).

\begin{table}[!htb]
\begin{center}
\caption{ The main background cross sections (in pb).}
\begin{tabular}{lcc}
\hline
\hline
Process      & $ \sigma_{LO}$ & $\sigma_{NLO} $  \\
\hline
${\rm t \bar t}$ & 505 & 830   \\
\hline
WW & 70 & 117  \\
\hline
WZ & 27 & 50  \\
\hline
ZZ & 11 & 16  \\
\hline
Wt & 30 & 62  \\
\hline
Z${\rm b \bar b}$ & 790  & 1580 \\
\hline
Z+jet, ckin(3) = 100 & 240  & 274 \\
\hline
DY2$\tau$ & 39600  &  \\
\hline
\hline
\end{tabular}
\label{Cross}
\end{center}
\end{table}

In the final analysis the events with the following isolation criteria 
for electrons were used:
$TrackIsolation < 1.0$, $CaloIsolation > 0.85$, $0.85 < EPratio < 2.0$, 
$HEratio <0.25$.
The same criteria for muons were the following: 
$TrackIsolation < 1.0$, $CaloIsolation > 0.50$, $EPratio < 0.20$, 
$HEratio > 0.70$. 
These numbers were adjusted by studying electron and muon tracks  in 
the process $ pp \rightarrow WW \rightarrow 2l$.

The selection cuts are the following:

\begin{itemize}
\item { 
cut on leptons: $p_T^{lept} > p_T^{lept, 0}$, $|\eta | < 2.4$, 
lepton isolation within ${\Delta}R < 0.3$ cone}
\end{itemize}

\begin{itemize}
\item { cut on missing transverse energy: $E_T^{miss} > E_T^{miss, 0}$. }
\end{itemize}

Where $p_T^{lept, 0}$ and $E_T^{miss, 0}$ are corresponding thresholds.

\subsection{            Trigger selection
}

The events are required to pass the Global Level 1 Trigger (L1) 
\cite{L1trigger} and the High Level Trigger (HLT) \cite{HLTtrigger}. 
The events have to pass at least one of the following triggers: 
single electron, double electron, single muon, or double muon. The used 
cut on leptons is more stringent than the cuts used in the HLT
for these triggers.

\section{
\label{Resca}
                        Use of the $e^{\pm}\mu^{\mp}~+~E^{miss}_T$
                        signature for the SUSY detection
}

The possibility to detect SUSY using the
CMS test points LM1 - LM9 \cite{LMPoints,PTDR2} are chosen for the detailed 
study of SUSY detection at CMS is investigated in this section. 
This study is based on the counting the expected
number of events for both the SM and the mSUGRA models.
The parameters of the CMS test points 
LM1 - LM9 are given in Table \ref{Points}.

\begin{table}[!htb]
\begin{center}
\caption{ The parameters of the CMS test points.}
\begin{tabular}{cccccc}
\hline
\hline
~Point~~~& $ m_0$ (GeV)  & $ m_{1/2}$(GeV) & $\tan{\beta}$ & $sign(\mu)$ & 
$A_0 $    \\
\hline
LM1 & 60 & 250 & 10 & + & 0 \\
LM2 & 185 & 350 & 35 & + & 0  \\
LM3 & 330 & 240 & 20 & + & 0 \\
LM4 & 210 & 285 & 10 & + & 0  \\
LM5 & 230 & 360 & 10 & + & 0  \\
LM6 & 85 & 400 & 10 & + & 0 \\
LM7 & 3000 & 230 & 10 & + & 0  \\
LM8 & 500 & 300 & 10 & + & -300 \\
LM9 & 1450 & 175 & 50 & + & 0 \\
\hline
\hline
\end{tabular}
\label{Points}
\end{center}
\end{table}

For the point LM1 (the point LM1 coincides with the post-WMAP point B 
\cite{Batt}) the distributions on $p_{T}^{lept}$, $E_{T}^{miss}$
 and $m_{inv}(e^+\mu^{-} + e ^{-} \mu^{+})$
for both background and signal events  are shown 
in Figs.\ref{figPt}-\ref{figMinv}.

It was found that the set of cuts with $p_T^{lept} >$ 20~GeV/$c$, 
$E_T^{miss} >$ 300~GeV is close to the optimal set (the highest significance 
with the best signal/background ratio). The 
results for the luminosity ${\cal L} = ~10~{fb}^{-1}$  
are presented in Table \ref{Bgsgemu}.

For other CMS SUSY test points 
LM2 - LM9 the results with the same set of cuts
are presented in Table \ref{Fv_res}. The 
significances definitions in this table are the following:
$S_{c12}=2(\sqrt{N_S+N_B}-\sqrt{N_B})$ \cite{Bity} and 
$S_{cL}~=~$ $\sqrt{2((N_S~+~N_B)~ln(1~+~\frac{N_S}{N_B})~-~N_S)}$ \cite{BartV}.

\begin{table}[!htb]
\begin{center}
\caption{ The expected number of events for backgrounds and for 
signal at the point LM1, ${\cal L} = 10~fb^{-1}$, 
$e^{\pm}\mu^{\mp}~+~E^{miss}_T$ signature.}
\begin{tabular}{lcc}
\hline
\hline
Process & 2 isolated leptons, $p_T^{lept} >$ 20~GeV/$c$ & $E^{miss}_T >$ 300~GeV \\
\hline
${\rm t \bar t}$ & 39679 & 79  \\
\hline
WW & 4356 & 4  \\
\hline
WZ   & 334 & 2 \\
\hline
ZZ & 38 &  0   \\
\hline
Wt & 3823 & 2  \\
\hline
Z${\rm b \bar b}$ & 315 & 0  \\
\hline
Z+jet & 1082 & 6  \\
\hline
DY2$\tau$ & 7564 & 0  \\
\hline
SM background & 57191 & 93  \\
\hline
LM1 Signal & 1054 & 329  \\
\hline
\hline
\end{tabular}
\label{Bgsgemu}
\end{center}
\end{table}

\begin{table}[!htb]
\begin{center}
\caption{ The number of signal events and significances for cut set 
with $p_T^{lept}~>$~20~GeV/$c$  and $E_{T}^{miss}~>$~300~GeV
for ${\cal L} = 10~fb^{-1}$, signature $e^{\pm}\mu^{\mp}~+~E^{miss}_T$.
The number of the SM background events $N_{B}$ = 93 (see Table \ref{Bgsgemu}).
}
 \begin{tabular}{crrr}
\hline
\hline
~~Point~~ & ~$N$ events & ~~~$S_{c12}$ & ~~~$S_{cL}$ \\
\hline
LM1 & 329 & 21.8 & 24.9  \\
LM2 &  94 &  8.1 &  8.6  \\
LM3 & 402 & 25.2 & 29.2  \\
LM4 & 301 & 20.4 & 23.1  \\
LM5 &  91 &  7.8 &  8.3  \\
LM6 & 222 & 16.2 & 18.0  \\
LM7 &  14 &  1.4 &  1.4  \\
LM8 & 234 & 16.9 & 18.8  \\
LM9 & 137 & 11.0 & 11.9  \\
\hline
\hline
\end{tabular}
\label{Fv_res}
\end{center}
\end{table}

It was found from the Tables \ref{Bgsgemu}-\ref{Fv_res} that for the point LM1
the significances are 
$S_{c12}/S_{cL} = 21.8/24.9$ for the $e^{\pm}\mu^{\mp}~+~E^{miss}_T$ signature.

The supersymmetry discovery potential for the mSUGRA model with 
$\tan\beta = 10$, $sign(\mu) = +$ ~~in the $(m_0, ~m_{1/2})$ plane 
(generalization of the point LM1) using the CMS fast simulation program 
$FAMOS\_1\_4\_0$ \cite{CMSSOFT} was also studied.
The CMS discovery potential contours for ${\cal L} = 1, 10$ and $30~fb^{-1}$ 
for the signature $e^{\pm}\mu^{\mp}~+~E^{miss}_T$ are shown in Fig.\ref{figemDP}.

\subsection{
                        The comparison of the FAMOS and the full 
                        simulation for the point LM1 
} 
 
The results obtained with the FAMOS code were compared with the 
full simulation results. 
The test point LM1 was used as the comparison object. 
The distributions on $p^{lept}_T$ and $E^{miss}_T$ of the LM1 signal 
for both FAMOS and ORCA are shown in Figs.\ref{figllPtOF},\ref{figllEtOF}.   

From the Figs. \ref{figllPtOF},\ref{figllEtOF} it is possible to conclude 
that the full and the fast simulation distributions on  $p^{lept}_T$ and 
$E^{miss}_T$ for the point LM1 are in reasonable agreement.

\section{
\label{Selefl} 
                        Search for lepton flavour violation
                        in the neutralino decays
}

In the MSSM the off-diagonal components of the slepton mass 
terms violate lepton flavour conservation. As it has been shown in 
Refs.\cite{Kras1994},\cite{Arka}  it is possible to look for lepton 
flavour violation 
at supercolliders through the production and decays of the sleptons.
For the LFV at the LHC, one of the most promising 
processes is the LFV decay of the second neutralino \cite{Agas},\cite{Hisa} 
$~\tilde{\chi}^0_2 \rightarrow \tilde{l}l 
 \rightarrow \tilde{\chi}^0_1~ll^{'}$,
where the non zero off-diagonal component of the slepton mass matrix leads 
to different lepton flavours in the final state. This mode is more 
sensitive to LFV compared to the direct Drell-Yan production of 
sleptons since the second neutralino $\tilde{\chi}^0_2$ can be copiously 
produced through the cascade decays of squarks and gluinos \cite{Kras1997}.
By using the above 
mode, LFV in $\tilde{e} ~-~ \tilde{\mu}$ mixing has been investigated 
in Refs.\cite{Agas},\cite{Hisa} at a parton model level for a toy detector. 
In this section the perspectives of LFV detection in CMS on the base of 
full simulation of both signal and background is studied. 
To be specific, the test point LM1 is studied. 
The signal of LFV $\tilde{\chi}^0_2$ 
decay is the two opposite-sign leptons ($e^+\mu^{-}$ or $e^{-}\mu^{+}$) in 
the final state with the characteristic edge structure. 
In the limit of lepton flavour conservation, the  process 
$ \tilde{\chi}^0_2 \rightarrow \tilde{l}l \rightarrow ll\tilde{\chi}^0_1$ 
has the edge structure for the distribution of lepton-pair invariant 
mass $m_{ll}$ and the edge mass $m^{max}_{inv}(l^+l^-)$ is expressed by 
slepton mass $m_{\tilde{l}}$ and neutralino masses 
$m_{\tilde{\chi}^0_{1,2}}$ as follows \cite{Abdu}:
\begin{equation}
(m^{max}_{inv}(l^+l^-))^2 = m^2_{\tilde{\chi}^0_2}(1 - \frac{m^2_{\tilde{l}}}
{m^2_{\tilde{\chi}^0_2  }})(1 - \frac{m^2_{\tilde{\chi}^0_1}}
{m^2_{\tilde{l}}  }   )
\end{equation}

The SUSY background for LFV comes from uncorrelated leptons 
from different squark or gluino decay chains. The SM background  
comes mainly from 
\begin{equation}
t\bar{t} \rightarrow b W b W \rightarrow bl bl^{'}\nu \nu^{'}.
\end{equation}
It should be stressed that the signature with $e^{\pm}\mu^{\mp}$ 
in the absence of LFV do not have the edge structure for the distribution 
on invariant mass 
$m_{inv}(e^{\pm}\mu^{\mp})$. As the result of LFV the edge structure 
for $e^{\pm}\mu^{\mp}$ events arises too. Therefore the signature of LFV is 
the existence of an edge structure for the $e^{\pm}\mu^{\mp}$ distribution.  
The rate for a flavour violating decay is determined by nonzero $\kappa$,
where

\begin{equation}
\kappa = \frac{Br(\tilde{\chi}^0_2 \rightarrow e^{\pm}\mu^{\mp}~\tilde{\chi}^0_1)}
{Br(\tilde{\chi}^0_2 \rightarrow  e^{+}e^{-}\tilde{\chi}^0_1, \mu^{+}\mu^{-}\tilde{\chi}^0_1)}
\end{equation}

In this paper the observability of LFV for the point LM1 was studied. 
For this purpose a special sample of events with 100\% LFV was prepared.
For $\kappa = 0.10$, $\kappa = 0.25$   the distributions of the number of 
$e^{\pm}\mu^{\mp}$ events on invariant mass $m_{inv}(e^{\pm}\mu^{\mp})$ 
(Fig.\ref{figemMinv})  clearly exhibit the edge 
structure, i.e. the existence of the lepton flavour violation in 
neutralino decays. 

It appears that for the point LM1 the use of an additional cut 
\begin{equation}
m_{inv}(e^{\pm}\mu^{\mp}) < 85~{\rm GeV}
\end{equation} 
reduces both SM and SUSY backgrounds and increases the  
discovery potential in the LFV search. In this case the cuts 
$p^{lept}_{T} >$ 40 ~GeV/$c$ and $E^{miss}_{T} >$ 200 ~GeV give the best result.
For instance, it was found that $N_{SMbg} = 19$ and 
$N_{SUSYbg} = N_{LM1_Signal} = 71$
for the $e^{\pm}\mu^{\mp}$ signature (see Table \ref{Bgsgemu} for 
comparison where $N_{SMbg} = 93$ and $N_{LM1_Signal} = 329$).

It was found that for the point LM1 in the assumption of exact knowledge of 
backgrounds (both SM and SUSY backgrounds) for the integral luminosity of 
${\cal L} = 10 fb^{-1}$ it would be possible to detect LFV 
at a $5 \sigma$ level in $\tilde{\chi}^0_2$ 
decays for $\kappa  \geq 0.04$.

\section{
\label{Insys}
                        Influence of the systematic uncertainties
                        on the signal significance 
}

In this Section the influence of systematic uncertainties on 
the value of signal significance in the case of SUSY detection is estimated.
The systematic uncertainties in the signal significance 
calculation include the experimental selection uncertainty of the background 
events, luminosity uncertainty, and the theoretically calculated uncertainties 
of the ${\rm t \bar t}$, WW and other backgrounds. 
Theoretically calculated 
uncertainty in background cross sections consists of the uncertainty related
with inexact knowledge of parton distribution functions (PDF) and higher order 
corrections to the NLO cross sections \cite{BartP}. 
There are several experimental 
uncertainties related with lepton identification, missing energy and luminosity. 
In accordance with Ref.\cite{Ptlept} the systematic error 
related with the lepton identification is 3\%, the
systematic error related with the missing energy is 2\% \cite{Etmiss}. 

It was found that the experimental uncertainties in the number of background
events $N_{B}$ related with the missing energy and the 
lepton identification lead to 10\% and 0.5\% uncertainties in the number
of the $N_{B}$ respectively.

The systematic uncertainty in the luminosity is 5\% \cite{Lumi}.
The total 5\% uncertainty in luminosity leads to 5\% uncertainty in the 
number of background events. 

In the studying signature the ${\rm t \bar t}$ background dominates. 
The PDF uncertainty of ${\rm t \bar t}$ cross section is equal to 5\% and 
the uncertainty due to unknown higher order corrections to the NLO background 
cross section is equal to 10\%. In the assumption that the systematic 
uncertainties are added quadratically it was found that the overall uncertainty 
in the number of background events is about 16\%. 

Following the prescriptions of the CMS PRS group the influence of the systematic 
uncertainties on signal significance using the program 
for calculations of significance S\_cP 
from Ref.\cite{ScP} was calculated. The results are shown in Table \ref{Uncert}.

\begin{table}[!htb]
\begin{center}
\caption{The dependence of signal significance $S_{c12}$ on 
background uncertainty for the used set of cuts and ${\cal L} = 10~fb^{-1}$
for signature $e^{\pm}\mu^{\mp}~+~E^{miss}_T$.} 
\begin{tabular}{crrrrr}
\hline
\hline  
~~~~Point~~~~ &    ~0\% &  ~~10\% &  ~~20\% \\
\hline
LM1 &   21.8 & 15.7 & 10.0  \\
LM2 &    8.1 &  5.8 &  3.7  \\
LM3 &   25.2 & 18.1 & 11.6  \\
LM4 &   20.4 & 14.7 &  9.4  \\
LM5 &    7.8 &  5.6 &  3.6  \\
LM6 &   16.2 & 11.7 &  7.5  \\
LM7 &    1.4 &  1.0 &  0.6  \\
LM8 &   16.9 & 12.1 &  7.8  \\
LM9 &   11.0 &  7.9 &  5.1  \\
\hline
\hline
\end{tabular}
\label{Uncert}
\end{center}
\end{table}

The discovery potential decreases with the increase of the background 
uncertainty but nevertheless the results are rather robust. In particular, 
it was found that for 0, 10, 20\%  background uncertainties it is 
possible to detect at the point LM1 lepton flavour violation in neutralino 
decays for $\kappa  \geq 0.04, 0.043, 0.051$ respectively.

\section{
\label{Conc}
                        Conclusion
}
 
In this paper the possibility to detect SUSY and lepton flavour violation  
at the LHC (CMS) using the signature $e^{\pm}\mu^{\mp}~+~E^{miss}_T$ 
was studied. This signature allows to discover both SUSY and LFV in 
the neutralino decays. 
It was found that for the CMS test points LM1 - LM9 for integral luminosity
${\cal L} = 10~fb^{-1}$ 
it is possible to discover SUSY for all LM points except for LM7. 
For $\tan\beta = 10,~ A = 0, ~sign(\mu) = + $ ~~the discovery contours for
${\cal L} =1,~10$ and $30~fb^{-1}$ in the ($m_0, ~m_{1/2}$) plane were 
determined. The possibility to look for lepton flavour violation in 
neutralino decays was also studied. It was found that for the point LM1 it is 
possible to detect lepton flavour violation provided the lepton flavour  
violating branching $Br(\tilde{\chi}^0_2 \rightarrow \mu^{\pm}e^{\mp}
\tilde{\chi}^0_1) \geq 0.04 Br(\tilde{\chi}^0_2 \rightarrow 
e^{+}e^{-}\tilde{\chi}^0_1, \mu^{+}\mu^{-} \tilde{\chi}^0_1)$
for ${\cal L} = 10~fb^{-1}$.  
The results for the significances are rather robust under the inclusion 
of reasonable systematic uncertainties. 

It should be stressed that the signature $e^{\pm}\mu^{\mp}~+~E^{miss}_T$
is less ``powerful'' from the point of view of SUSY discovery that say
the signature $(n~ \geq ~3 ~jets) + E^{miss}_T$ \cite{Abdu}, however it is 
important to detect ``new physics'' using simultaneously different 
signatures that increases the credibility of
the results. 

\vspace{1cm}

\section{
                        Acknowledgments
}

The authors would like to thank D.Denegri and L.Rurua for careful reading 
of the manuscript, useful discussions and suggestions concerning the note.\\
The authors also would like to express their thanks to L.Pape and M.Spiropulu for
interest and very useful comments.\\ 
The authors would like to thank S.Slabospitsky for useful discussions.\\
\\
This work  has been supported by RFFI grant No 04-02-16020.

\newpage

\begin{figure}
\includegraphics{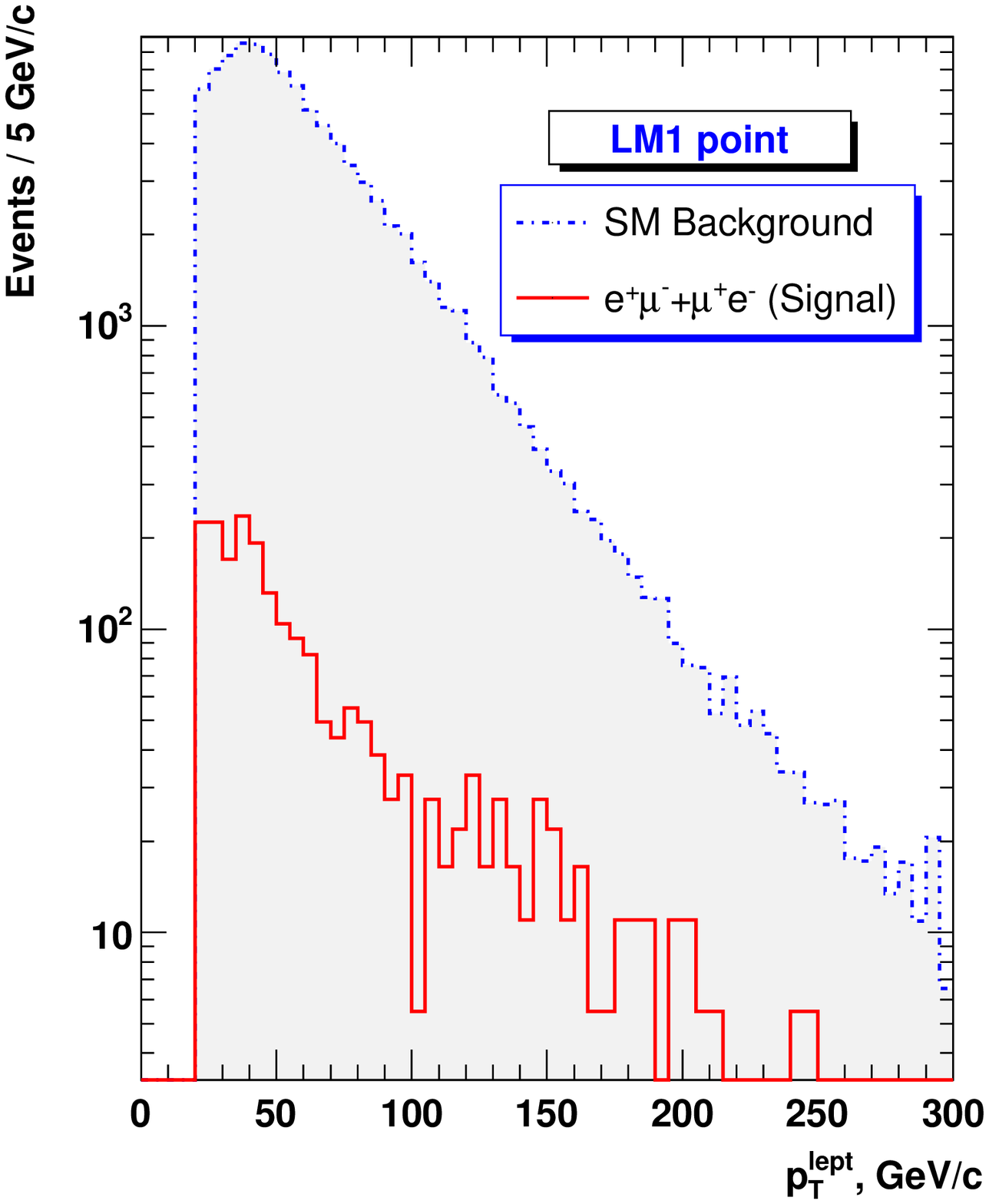} 
\caption{The $p^{lept}_T$ distribution after selection of two 
      isolated leptons with $p^{lept}_T >$ 20~GeV/$c$. The both
      leptons are plotted.}
    \label{figPt} 
\end{figure}
 
\begin{figure}
\includegraphics{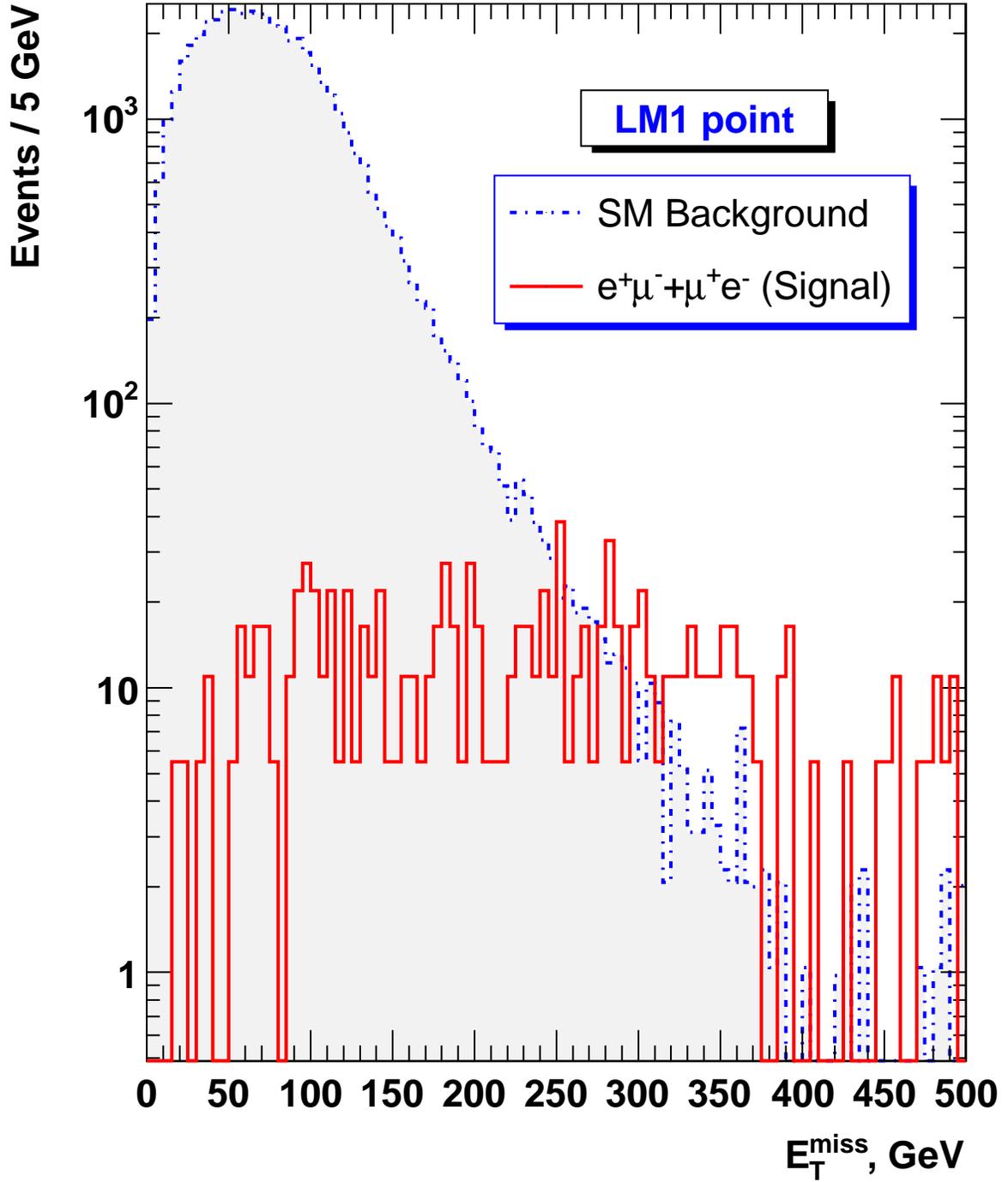} 
\caption{The $E_T^{miss}$ distribution after selection of two isolated 
      leptons with $p_T^{lept} >$ 20~GeV/$c$.}
   \label{figEt} 
\end{figure}
 
\begin{figure}
\includegraphics{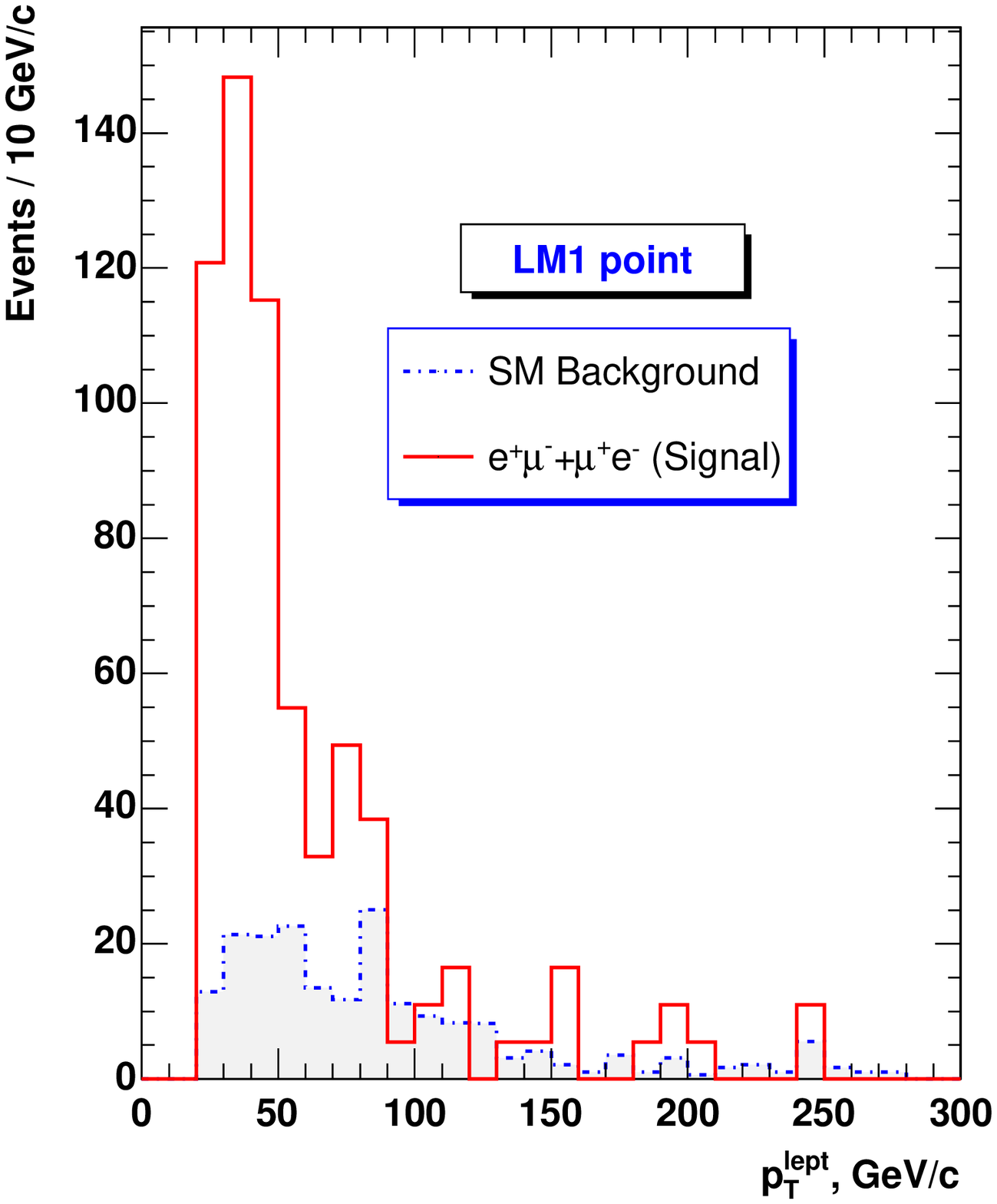} 
\caption{The $p^{lept}_T $ distribution 
      after selection of two isolated leptons with  $p^{lept}_T >$ 20~GeV/$c$ 
      and $E_T^{miss} >$ 300~GeV.}
   \label{figPt2} 
\end{figure}
 
\begin{figure}
\includegraphics{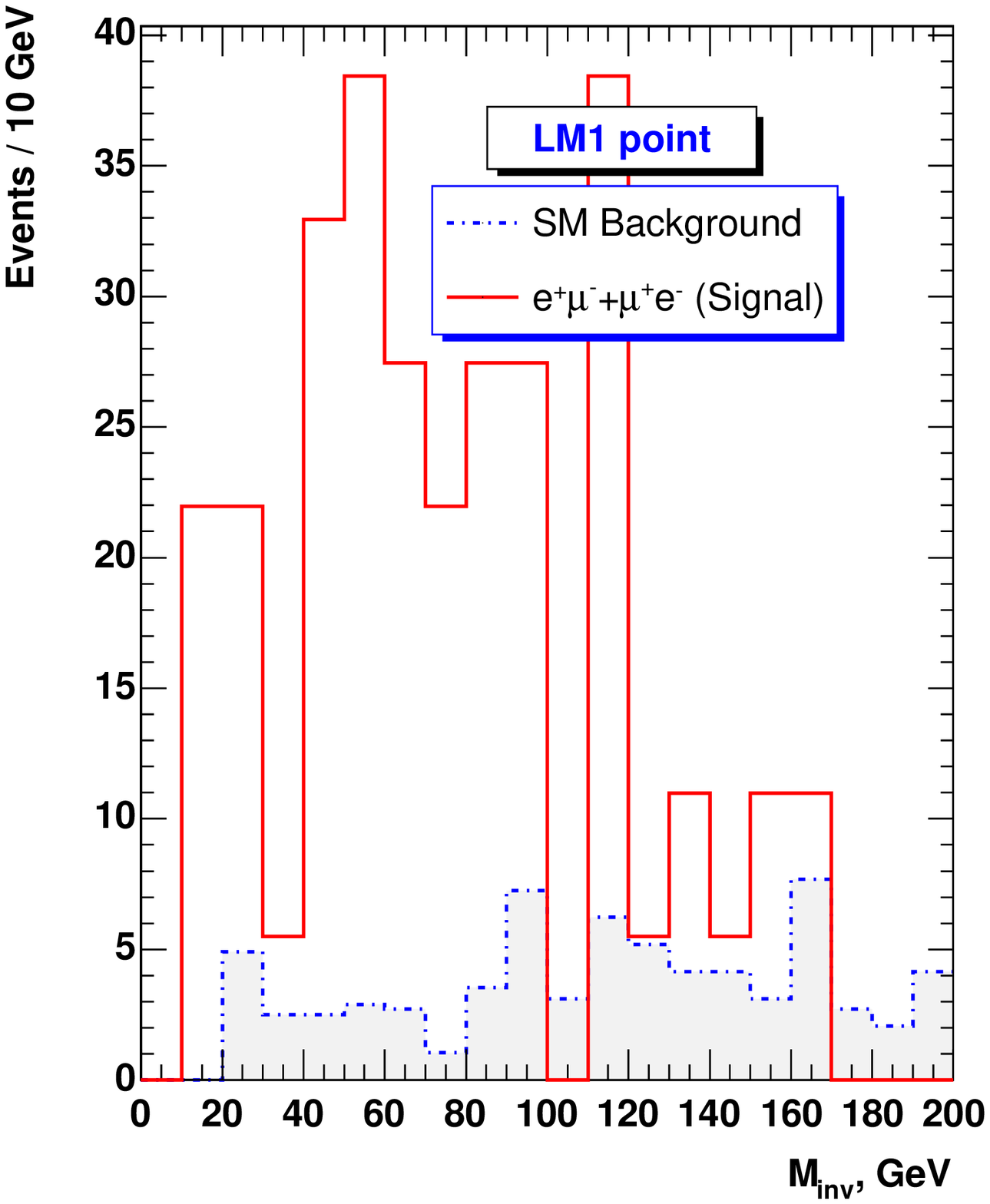} 
\caption{The invariant two lepton mass distribution 
      after selection of two isolated leptons with $p_T^{lept} >$ 20~GeV/$c$ and 
      $E^{miss}_T >$ 300~GeV.}
   \label{figMinv} 
\end{figure}
 
\begin{figure}
\includegraphics[width=14.0cm]{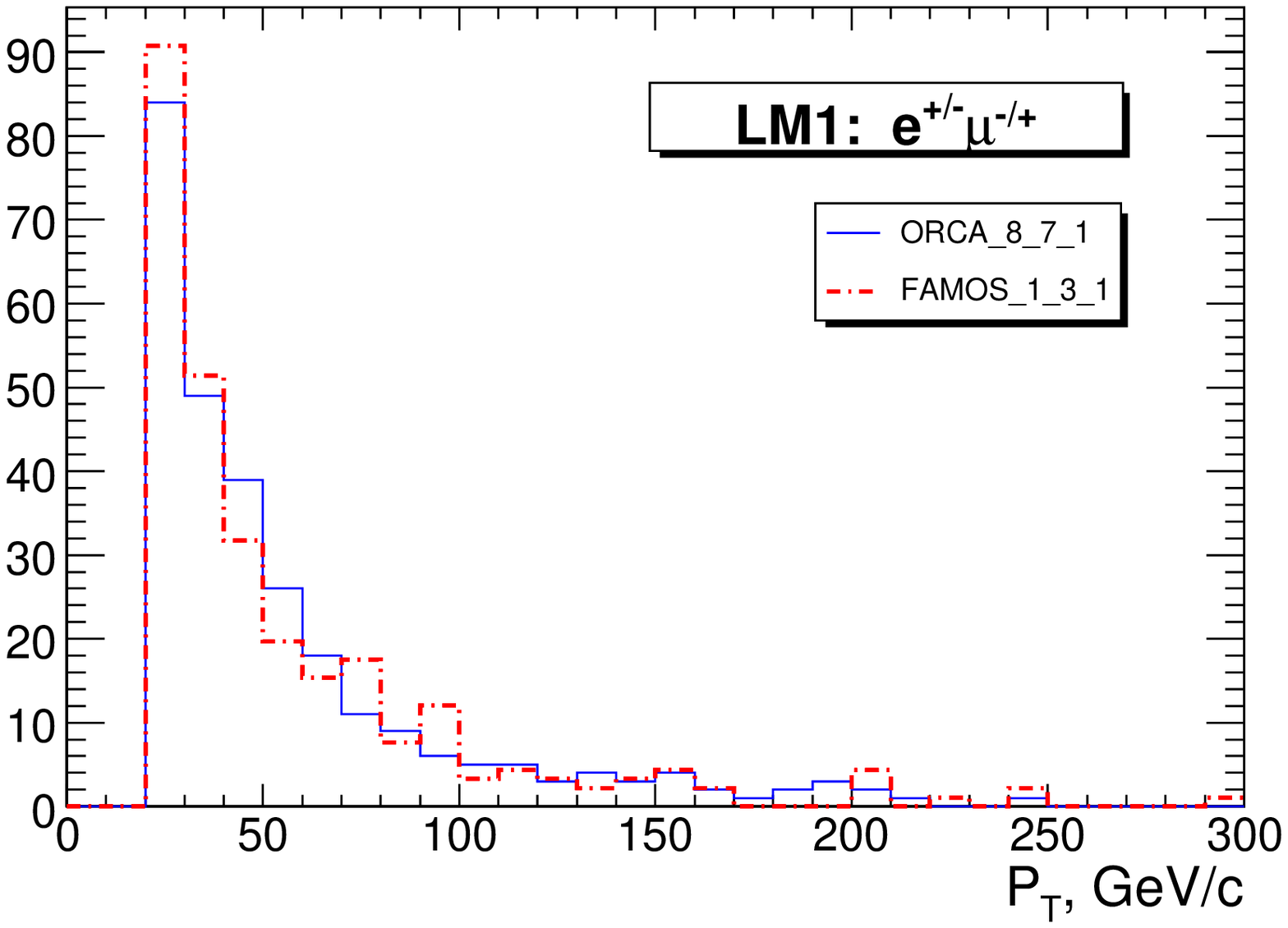} 
\caption{The $p_T^{lept}$ distribution 
      after selection of two isolated leptons with $p_T^{lept} >$ 20~GeV/$c$ 
      for ORCA and FAMOS. The plots are normalized to the numbers of 
      dileptons.}
   \label{figllPtOF} 
\end{figure}
 
\begin{figure}
\includegraphics[width=14.0cm]{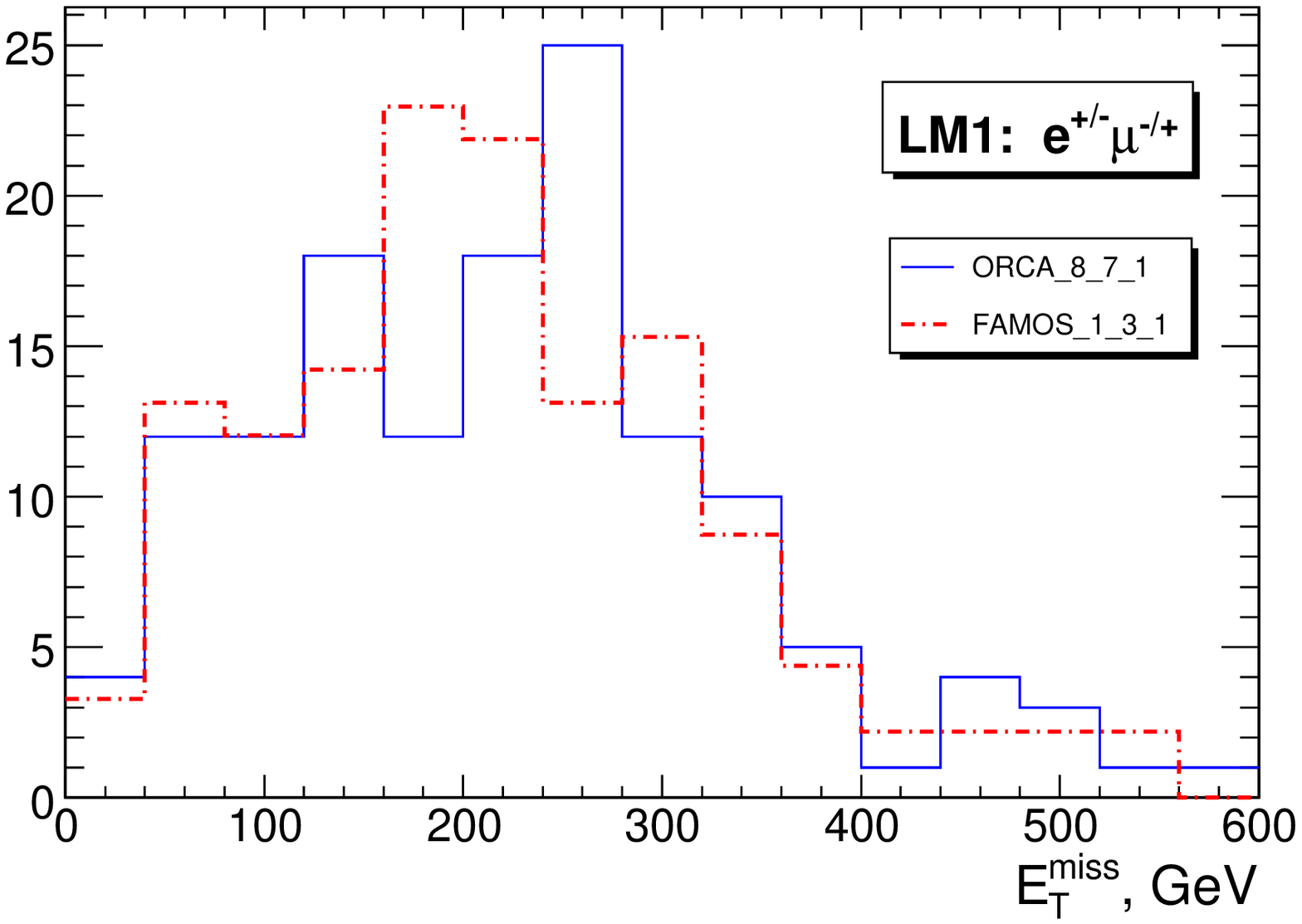} 
\caption{ The $E_T^{miss}$ distribution after selection of two isolated 
      leptons with $p_T^{lept} >$ 20~GeV/$c$ for ORCA and FAMOS. 
      The plots are normalized to the numbers of dileptons.}
   \label{figllEtOF} 
\end{figure}
 
\begin{figure}
\hbox{\resizebox{8cm}{!}{\includegraphics{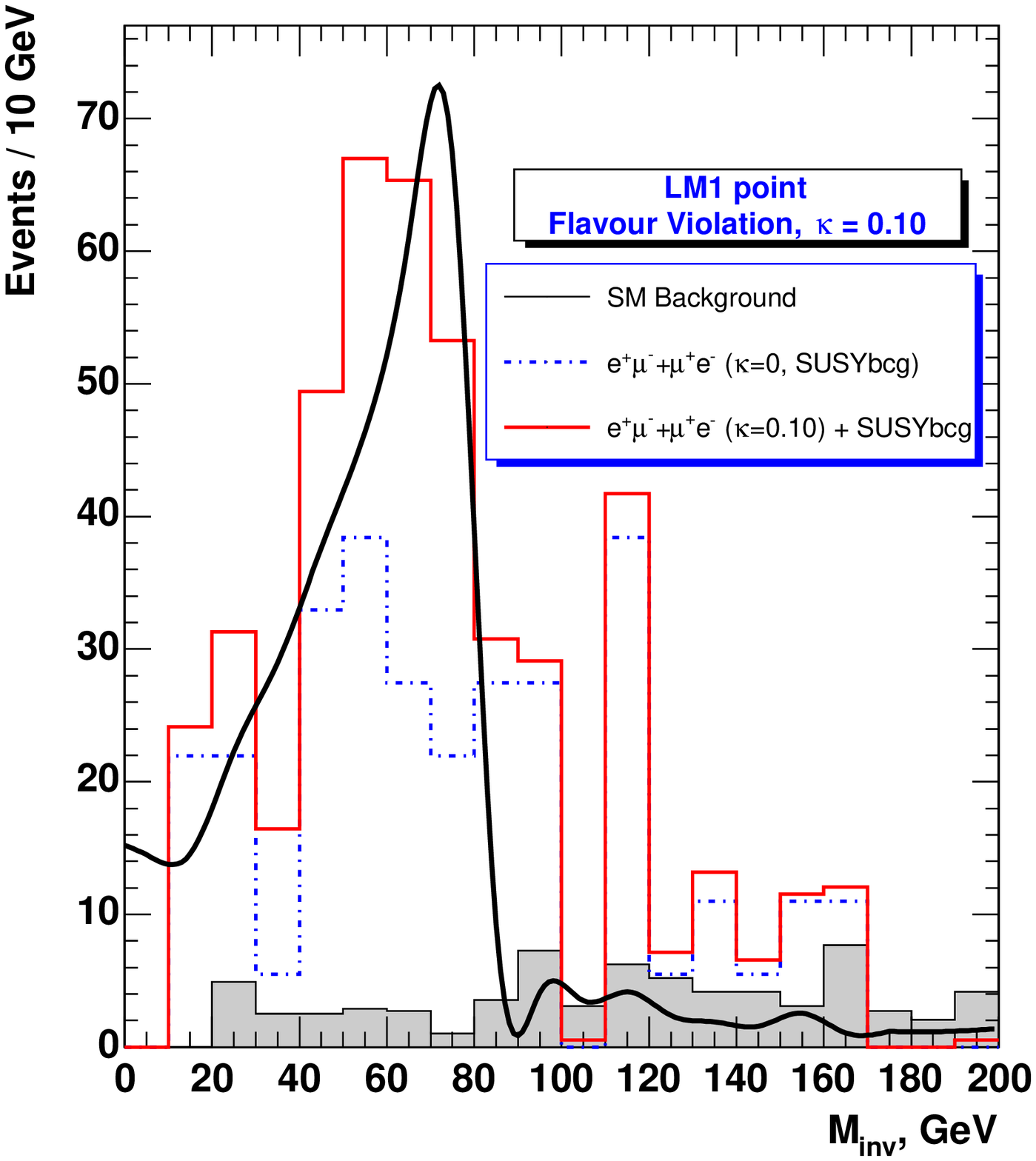}}
      \resizebox{8cm}{!}{\includegraphics{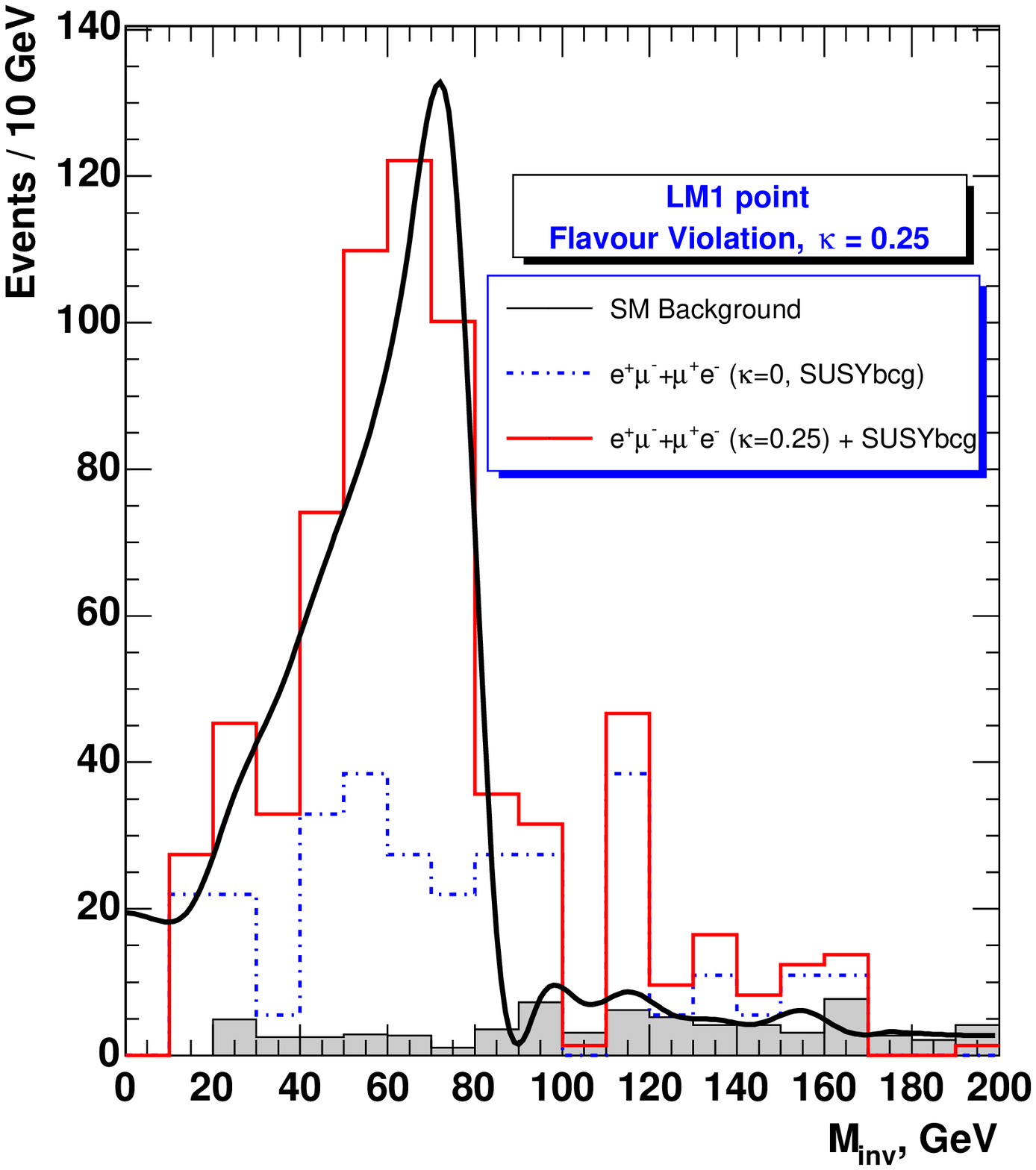}}}
\caption{ The distribution of two lepton invariant mass after selection 
    of two isolated different-flavour leptons with $p_T^{lept} >$ 20~GeV/$c$ 
    and $E^{miss}_T >$ 300~GeV for flavour violation parameter $\kappa = 0.10$
    and $\kappa = 0.25$. Superimposed curves are fits to the 
    invariant mass distribution for 100\% LFV.}
   \label{figemMinv} 
\end{figure}

\begin{figure}
\includegraphics[width=14.0cm]{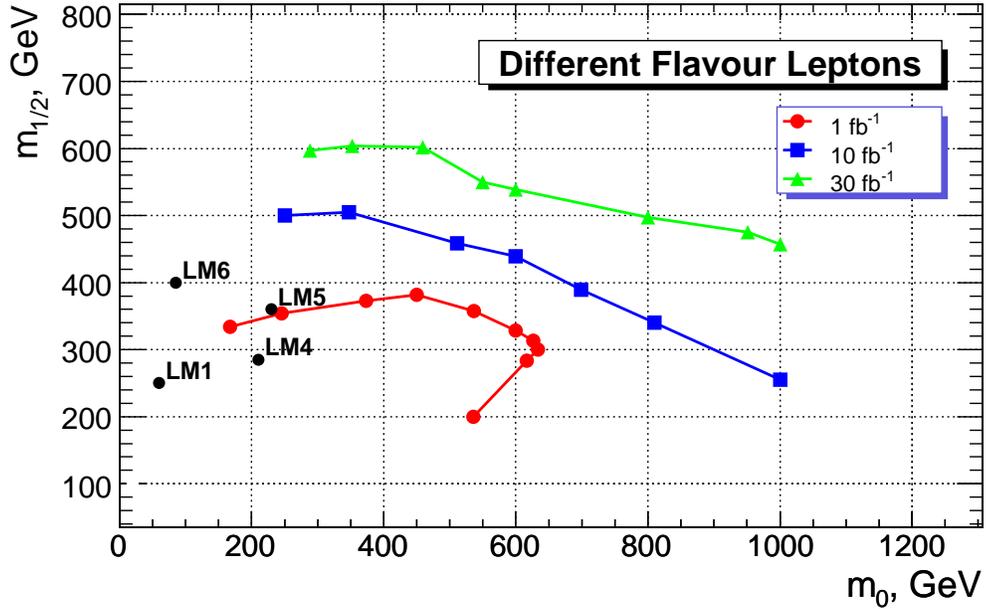} 
\caption{$e^+\mu^{-} ~+ ~e^{-}\mu^{+}$ discovery plot 
      for $\tan{\beta} = 10$, $sign (\mu) = +$, $ A = 0$.
      Selected two isolated leptons with  $p_T^{lept} >$ 20~GeV/$c$ and 
      $E^{miss}_T >$ 300~GeV. Calculations are made for $S_{c12}$.}
   \label{figemDP} 
\end{figure}
 
\end{document}